# LA-UR-26-21645
**Approved for public release; distribution is unlimited.**

**Title:** Ground Effects of the 2024 Mother's Day Superstorm: A Multi-source Observational Analysis

**Author(s):** Chen, Yue
Kim, Kyoung Ho
Morley, Steven Karl
Woodroffe, Jesse Richard

**Intended for:** Report

**Issued:** 2026-03-16 (rev.1)

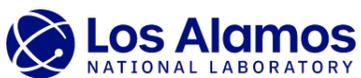 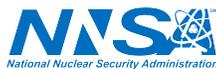 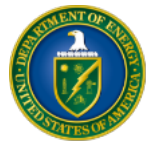

# Ground Effects of the 2024 Mother's Day Superstorm: A Multi-source Observational Analysis

Yue Chen (cheny@lanl.gov)[1], Kyoung Ho Kim[1], Steven K. Morley[1], Jesse R. Woodroffe[1]

[1] ISR-1 Group, Los Alamos National Laboratory

March 1, 2026

**Abstract.** This report presents a brief review of the 2024 Mother's Day superstorm and its impacts on the near-Earth space environment and the ground-level effects, with emphasis on the latter. Drawing upon observations from multiple sources. we qualitatively illustrate how intense space weather disturbances can generate strong geoelectric fields and drive pronounced geomagnetically induced currents, as reported at numerous locations, particularly within the New Zealand power grids.

1. Timeline of Key Space Weather Events

The 2024 Mother's Day superstorm is, to date, the most intense geomagnetic storm of Solar Cycle 25, producing a list of significant space and ground disturbances to the Earth. All these disturbances originated from the long-lasting NOAA Active Region (AR) 13664, located in the southern hemisphere of the solar photosphere, as the representative snapshot shown in Figure 1. AR 13664 was visible for two weeks from May 1-14 in 2024 and later reemerged with a new designation during the subsequent solar rotation period. Over this two-week interval, this active region generated dozens of solar flares, including multiple X-class events beginning on May 8 as summarized in Figure 2, some of which were accompanied by Earth-directed coronal mass ejections (CMEs).

Specifically, between May 8-9, at least five Earth-bounding CMEs sourced from AR 13664 were observed, with high speeds of ~1000 km/s. CCMC simulations[1] suggest that these CMEs, after first sweeping past the Mercury, reached the Earth in the afternoon of May 10 (Figure 3), triggering the Mother's Day super geomagnetic storm from May 10-12 and causing significant geoeffective impacts on near-Earth space plasma, the geomagnetic field and ground-based power grids. After this, these CMEs continued propagating through the heliosphere until reaching the Heliopause[2].

---

[1] CCMC CME scoreboard at https://kauai.ccmc.gsfc.nasa.gov/CMEscoreboard/prediction/detail/3163

In Earth's vicinity, after travelling ~2 days, the combined interplanetary (IP) shocks driven by the multiple CMEs first arrived at the L1 point at 1637 UTC on May 10, as observed in-situ by the upstream solar wind monitor – the Wind satellite. Half an hour later at 1705 UTC, the forward shock front collided with the Earth's bow shock nose, showing impacts from high speed (>700 km/s), elevated temperature, enhanced pressure (Psw>~50 nPa), and strongly magnetized solar wind (Figure 4A-D). The collision immediately triggered geomagnetic disturbances, marked by the AE index exceeding 1000 nT from 1707 UTC and a sudden storm commencement (SSC) in Sym-H > 60 nT at the same time (Figure 4 E & G). Subsequently, several prolonged intervals of southward IMF embedded within the complex shock structure facilitated intermittent dayside magnetic reconnections and subsequent substorm injections at nightside as indicated by high AE values often well above 1000 nT (Figure 4G and H). These repetitive substorm injections conveyed substantial new plasma deeply into the inner magnetosphere, significantly intensifying the ring currents and further disturbing the geomagnetic field. The resulting super geomagnetic storm reached a minimum Sym-H value of nearly -500 nT at around 0210 UTC on May 11, followed by a prolonged recovery phase lasting days (Figure 4E). During the course, the Asy-H index exhibited multiple peaks, with maximum values beyond 800 nT during the storm main phase and additional smaller peaks (still > 300 nT) in the early recover phase prior to 0440 UTC on May 12 (i.e., Time 10 in Figure 4F). Between Time 1 (the arrival of the forward shock front) and Time 10 (the arrival of reverse shock front), substantial substorm activities were accompanied with intensified widespread aurora emissions, often observable from ground-based All-sky Imagers[3] (not shown), consistent with intensified field-line current systems.

At ground level, the geoelectric field induced by variations in the geomagnetic field (Figure 4I) was clearly observed at Station BOU, with several pronounced peaks (e.g., at Time 3) corresponding to Sym-H peaks. Consequently, following the arrival of forward shock at 1705 UTC, numerous NERC locations across the USA recorded enhanced geomagnetically induced currents (GICs). For example, NERC Device 10692 observing its first GIC peak of 4.9A (LT~11am) at SSC as shown in Figure 4J. As illustrated in Figure 4, the interval of continuous enhanced GICs at Derice 10692—from UT 1700 to UT 5300—closely coincides with the interval of large variations in geomagnetic indices as well as the interval between the arrival of forward shock (Time 1) and the passage of the reverse shock front (Time 10). In other words, this individual storm clearly demonstrates how space weather disturbances can propagate through space and produce significant impacts on ground infrastructure.

---

[2] Portions of energetic neutral atoms (ENA) associated with this event are expected to be reflected inside the interstellar space and travel back toward the Sun, potentially being detectable to sensitive ENA instruments aboard the IMAP satellite ~ two years later (i.e., 2026-2027), based on the previous ENA studies (e.g., D. J. McComas *et al.*, 2018, doi: 10.3847/2041-8213/aab611).

[3] UCalgary Space Remote Sensing Open Science Platform https://data.phys.ucalgary.ca/

In addition to the disturbed EM fields, energetic particle populations inside near-Earth space also underwent large intensifications during this superstorm[4]. Solar energetic protons (SEPs) exhibited multiple intensifications during the interval: the first small one began at 1335 UTC on May 10, prior to the arrival of the first IP shock, and peaked at 1745 UTC; the second high SEP peak occurred in the morning of May 11 during the main phase of the storm, leading to the most recent ground-level neutron event (GLE) 74[5]; and the third major SEP peak was in the early morning of May 14. Meanwhile, MeV electrons in the outer radiation belt first experienced a dropout during the storm main phase on May 10-11, and their recovery started around midday on May 11, coinciding with the early recovery phase of the storm. MeV electron flux levels eventually reached ~10 times of the pre-storm values by the late storm recovery phase on May 15.

As shown in Figure 4, the period of continuous major GIC events concluded by May 12, as solar wind conditions became more stable (though still at elevated speeds) and geomagnetic indices recovered steadily with minimum sudden variations. In contrast, elevated levels of energetic charged particles inside the near-Earth space persisted significantly longer, with noticeable effects continuing beyond May 15. These observations underscore that GICs are closely tied to the rapid reconfigurations of near-Earth space current systems, particularly in response to upstream solar wind discontinuities (e.g., shock fronts) and substorm plasma injections, while the dynamics of energetic particles in near-Earth space are governed by distinct physical processes operating on different timescales. Table 1 summarizes and compares the timelines of the key events discussed above.

## 2.  Comparison of Global GIC Observations

Figure 4 is further examined to investigate the potential cause-and-effect relationships among upstream solar wind disturbances, global geomagnetic indices, and NERC GIC measurements[6]. Determined from Panels I and J, several prominent geoelectric field and GIC peaks are marked with red vertical lines to establish their temporal correspondence (or lack thereof) with features in the OMIN data. For example, Times 1, 3, 5, 6, 7, 9 and 11 appear to be associated with abrupt changes in Psw, while Times 2, 3, 4, 6, 8, 9 and 10 coincide with sudden variations in the ASY-H and/or AE indices that could have strong local time (LT) effects. The GIC time series in Panel J is subsequently used as a reference in further analyses.

---

[4] From in-situ energetic particle observations made by GPS CXD sensors: https://www.ngdc.noaa.gov/stp/space-weather/satellite-data/satellite-systems/lanl_gps/version_v1.10r1/

[5] GLE list can be accessed at https://gle.oulu.fi/

[6] NERC data website: https://eroportal.nerc.net/gmd-data-home/search-gmd-data/

Figures 5 -6 present GIC observations/simulations from multiple locations. Figure 5 compares observations from five NERC sites in the USA to assess geographic variabilities: Devices 10692 locates near the center of the selected region, Devices 10191, 10692 and 10202 are distributed across different latitudes but share similar geographic longitude (~W90.5°), and Devices 10103, 10692 and 10428 are distributed across different longitudes (i.e., different LTs) but have similar geographic latitude values (~N37°). Figure 6 further compares GICs from the USA with previously reported GICs from New Zealand and the UK.

In Figure 5, we compare the temporal evolution of GICs at locations across the USA continent to distinguish global effects from possible local influences associated with different space current systems. The normalized GIC from Device 10692 is overplotted in green as a reference for comparison with measurements from other Devices/locations. Across all panels, the overall temporal variations generally follow the reference curve, although noticeable differences are evident. For example, both the reference and other sites in Panels B and C show their highest GIC peaks at Time 3, whereas the sites in Panels A and D reach their highest GIC peak at or near Time 4. Also, the peak heights at the SSC vary by location. For all NERC locations, within one minute of the shock arrival at 1705 UTC on May 10, all measured GICs show enhancements: Device 10191 recorded GIC=5.5A (LT at ~11 am), 10692 GIC=4.9A (LT at ~11am), 10202 GIC=3.9 A (LT ~11am), 10103 GIC=2.1 A (LT~9am), and 10428 GIC=12.6A (LT~12 noon). These difference values could reflect the latitude dependence as noted by Piersanti et al. (2025), and/or also may be due to different local ground conductivity. Another noticeable feature is that the black curve in Panel C has much more pronounced variations during the two-hour interval UT 17-19 after Time 1 and the interval UT 31-33, which are not seen at other locations. Variations in the latter interval may be caused by substorm injections, as indicated by concurrent AE spikes, when the Device in Panel C was near local midnight. Overall, the period of sustained GIC activities—from UT 1700 to UT 5300 for all locations in Figure 5—is confirmed to coincide with the interval of large variations in geomagnetic indices as well as with the arrival (Time 1) and passage (Time 10) of the IP shock fronts.

Figure 6 further examines location dependence of GICs by expanding to additional global locations. Based on literature reports, we compile global GIC observations from other continents, as shown in Figure 6. Here Panel A replots the normalized GIC from Device 10692 in the USA, Panel B shows GICs observed in New Zealand (adapted from the Transpower Report, 2024), and Panel C presents simulated GIC magnitudes at UK locations as reported by Lawrence et al. (2025). All three panels are aligned by time for comparison. It is evident that, for most Times such as 1, 2, 3, 5, 6, and 7 (possible also 10 and 11 with fewer observations available), all regions exhibit significant GIC peaks, albeit with different relative amplitudes. In contrast, at other Times such as 4, 8 and 9, enhancements appear to be more regionally confined; for example, UK sites show limited GIC responses at Times 4 and 8, while NERC GICs are relatively small at Time 9. Bases on these, we provisionally classify the marked Times into two categories at the bottom of Figure 6: "G" denotes the Times with significant global GIC enhancements, and "L" denotes intervals characterized by primarily localized GIC responses. The physical drivers underlying these two categories warrant future investigations through incorporation of additional observations and/or model simulations.

Ideally, simultaneous measurements of both GIC and induced electric field would be available to minimize uncertainties in interpreting their relationship; however, such coincidence datasets remain rare due to the limited number of operational geoelectric monitoring stations, particularly in North America.  In this study, we utilize measurements of GIC and induced geoelectric field from two neighboring locations near Boulder, USA, as shown in Figure 7. The resemblance is evident between the GIC time series in Panel A and the horizontal components of induced E field at BOU[7] in Panel B, particularly for the north-south component (in black) at most specified Times. Nevertheless, there are moments when the east-west E component may play a dominant role. For example, the GIC value at Time 2 is smaller than that at Time 3 despite the higher E north-south component at Time 2, which suggests that the east-west E component contributes significantly at one or both times. (Notably, in Panel B, the "bay" feature in the east-west component on May 12 is not accompanied by neither corresponding variations in the north-south component nor in the GIC, implying it could arise from factors other than large-scale space current systems.)  When available, these coincidental GIC and E field datasets would provide valuable benchmarks for validating the accuracy of local ground conductivity models and improving the accuracy of GIC simulations.

For the six NERC device locations, their maximum absolute GIC values during the whole storm were: 18.5A for Device 10191, 18.3A for Device 10692, 9,38A for Device 10202, 10.5A for Device 10103, 42.7A for Device 10428, and 10.5A for Device 10418.  The first three Devices, separated in latitudes but sharing similar longitudes, follow the expected trend of decreasing GIC towards lower latitudes; however, additional factors—including LT dependence of related space current systems and varying local ground conductivities—likely contribute to the substantially different GIC values among the other Devices. In addition, throughout the whole interval, the induced geoelectric components measured at USGS BOU station remained below 1100 mV/km, keeping

---

[7] E-field data from BOU station are available at
https://geomag.usgs.gov/plots/?channels=U&channels=H&channels=X&channels=D&channels=E-E&channels=E-N&channels=Y&endTime=2024-05-13T00:00:00.000Z&startTime=2024-05-10T00:00:00.000Z&timeRangeType=custom&stations=BOU

the |E| values well under the TPL-007[8] local benchmark of ~5400 mV/km. In comparison, New Zealand reported a maximum induced current of 48 A at South Dunedin substation (SDN) transformer along with multiple grid emergencies (Transpower Report, 2024), while maximum GIC in UK exceeds 60 A as the simulations shown in Figure 6C. Beyond the latitude and local time effects, these pronounced differences in maximum GIC amplitudes may also reflect the regional variations in ground conductivity as well as differences in specific power grid designs and network topology. Again, like the work by Piersanti et al. [2025], quantitively determinations of induced geoelectric field and GIC responses based on local geomagnetic disturbances will be presented in a separate report.

## 3. Summary and Conclusions

In this report, we qualitatively examine the cause and effects of the 2024 Mother's Day superstorm on Earth. By integrating observations from multiple sources, we trace this significant event from its solar origin, through the propagation of CME-driven disturbances across the heliosphere, and eventually to their impacts on the near-Earth space radiation environment, the geomagnetic field, the induced geoelectric field at the ground, and ultimately the GICs measured at diverse global locations. As summarized in Table 1, during this superstorm, the continuous GIC activities coincided with the interval of the major IP shock structure passing through the Earth, which occurred ~60 hours after the first related CME being observed. The GIC interval started with the SSC and persisted for totally ~30 hours, spanning over the storm main phase and early recovery phase. It is also shown that the GIC responses during this superstorm are largely global, although the induced current intensities varied substantially due to regional factors (e.g., local ground conductivity, geographic/magnetic latitude, local time and specific grid configuration). This report serves as a pilot case study illustrating the close coupling between solar eruptions, space weather disturbances, and their ground-level impacts. Further quantitative modeling efforts based on this report are expected to help reveal gaps in our current physical understanding and improve forecasting capabilities for future extreme space weather events.

---

[8] Available from https://www.nerc.com/standards/reliability-standards/tpl/tpl-007-4

**Figures**

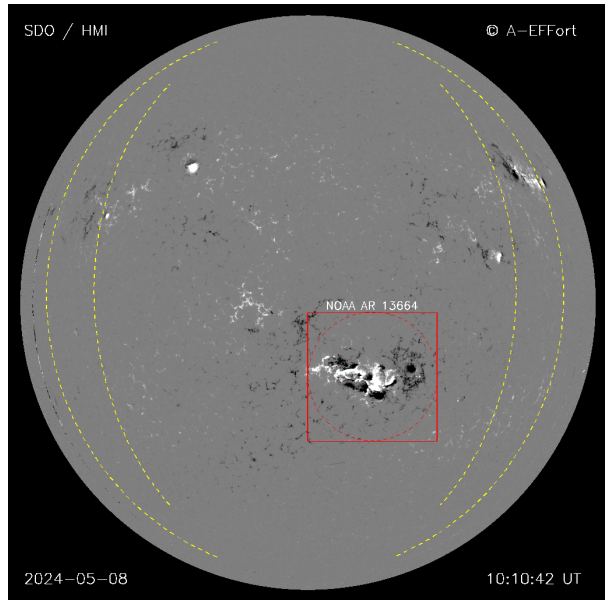

**Figure 1**. One image of AR 13664 taken on May 08, 2024, by the Helioseismic and Magnetic Imager (HMI) aboard the Solar Dynamics Observatory (SDO). Image credit: NASA/SDO and the HMI science team.

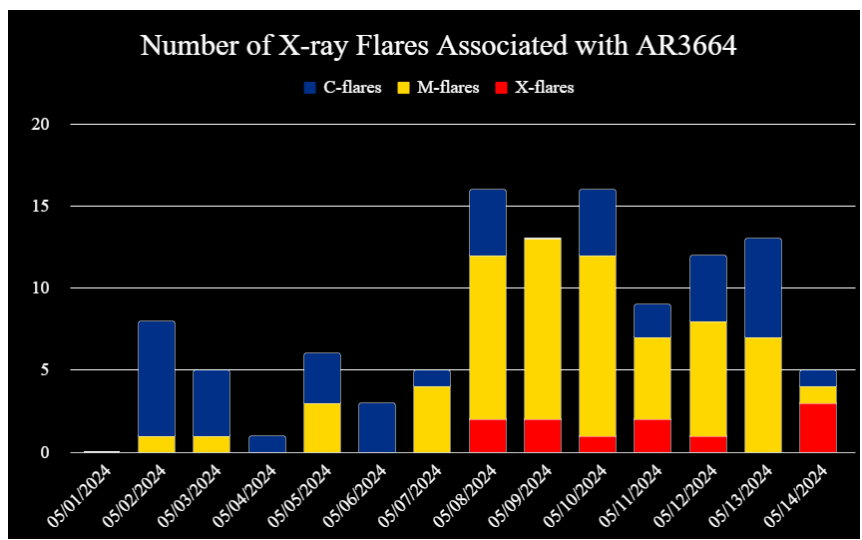

**Figure 2**. The number and classification of X-ray flares originated from AR 13664. Image credit: NOAA/SWPC.

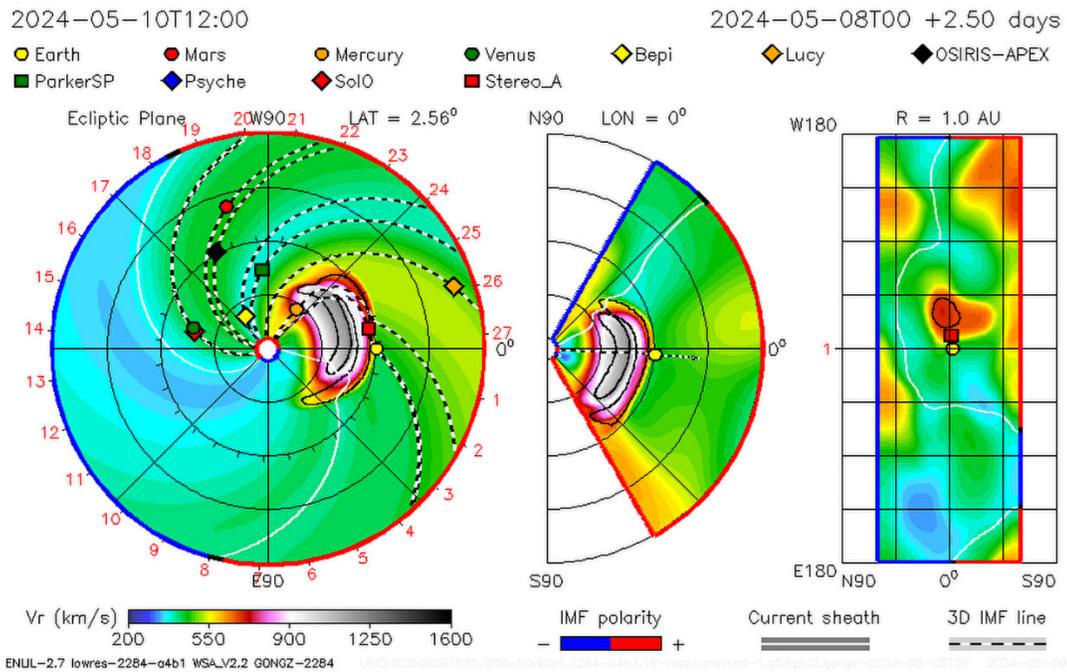

**Figure 3**. Snapshot of CME-driven shock fronts travelling toward Earth through the Heliosphere. Shown here are ENLIL-simulated radial velocity distributions in 3D space at T12:00:00Z on May 10, 2024, when the interplanetary (IP) shocks were still several hours from Earth. Credit: CCMC CME scoreboard at https://kauai.ccmc.gsfc.nasa.gov/CMEscoreboard/prediction/detail/3163.

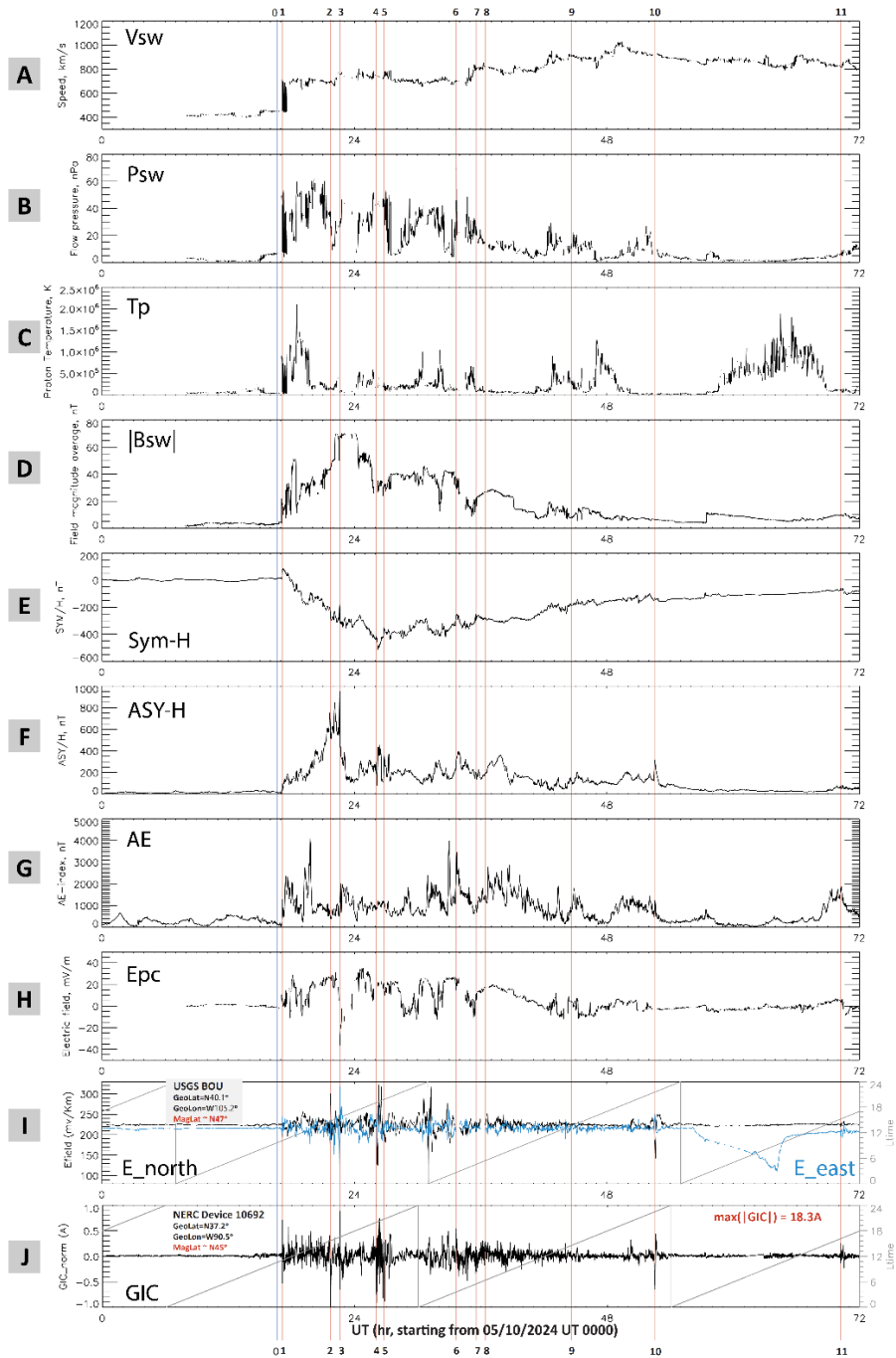

**Figure 4**. OMNI and ground-based data over the three-day interval May 10-12, 2024. From top to bottom: solar wind speed (Vsw), flow pressure (Psw), proton temperature (Tp), IMF magnitude (|Bsw|), Sym-H index, Asy-H index, electric field over the north polar cap (Epc), horizontal geoelectric field components at BOU station, and GIC—normalized to the interval maximum of 18.3 A—observed by device 10692 located at GEO latitude of 37.5° and longitude of 90.5°. In panels I and J, local time traces of station locations are overplotted in gray. Vertical lines mark key moments: "0" indicates the arrival of IP shocks at L1, and other lines denote GIC peaks identified at several locations, including the maximum |GIC| recorded by Device 10692 at Time 2.

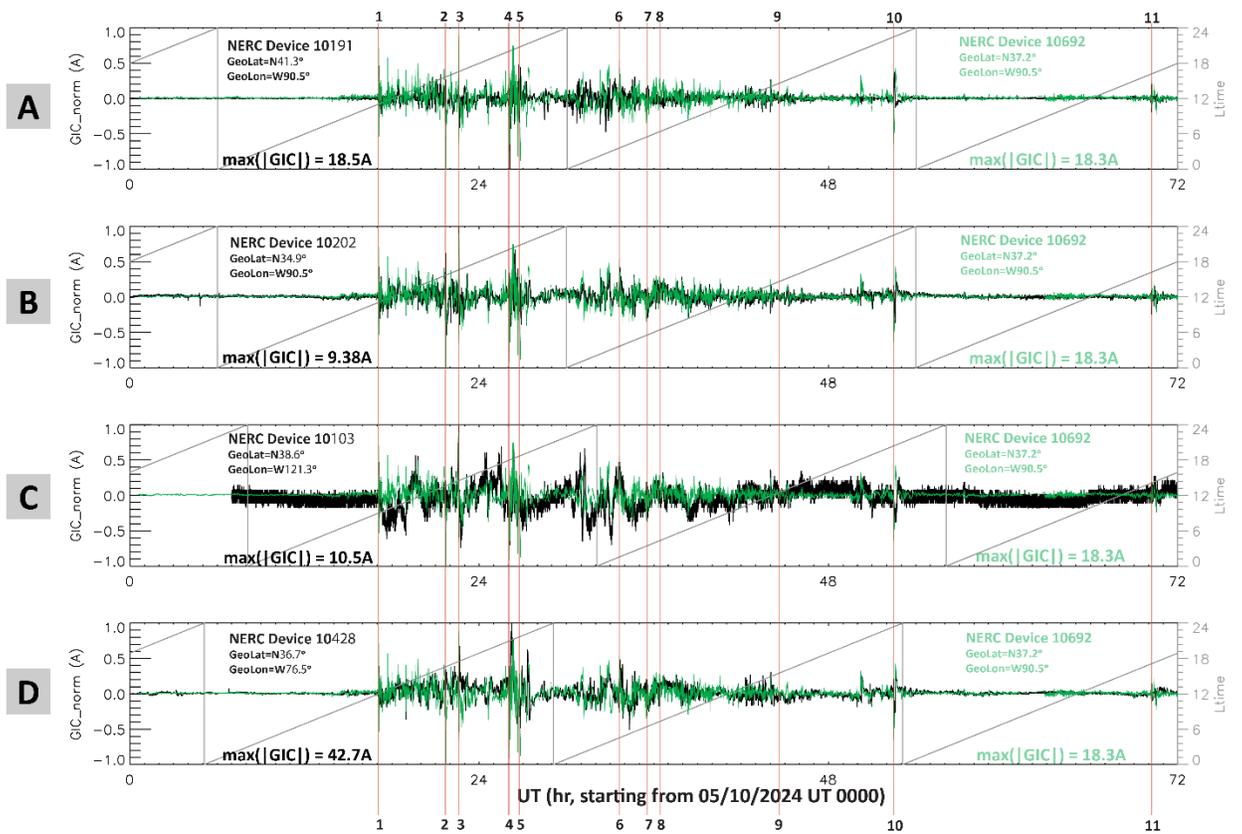

**Figure 5**. Normalized GIC observations from four USA locations compared to the reference Device 10692 (green curve) over the same three-day interval. In Panel A, the black curve represents a location with ~4° higher latitude but same longitude; Panel B shows ~2° lower latitude at the same longitude; Panel C shows similar latitude but ~30° larger longitude (i.e., ~ 2 hr difference in local time), and Panel D shows ~15° smaller longitude (i.e., ~ 1 hr difference in local time) at a similar latitude. The maximum |GIC| values of individual locations are indicated in the panels.

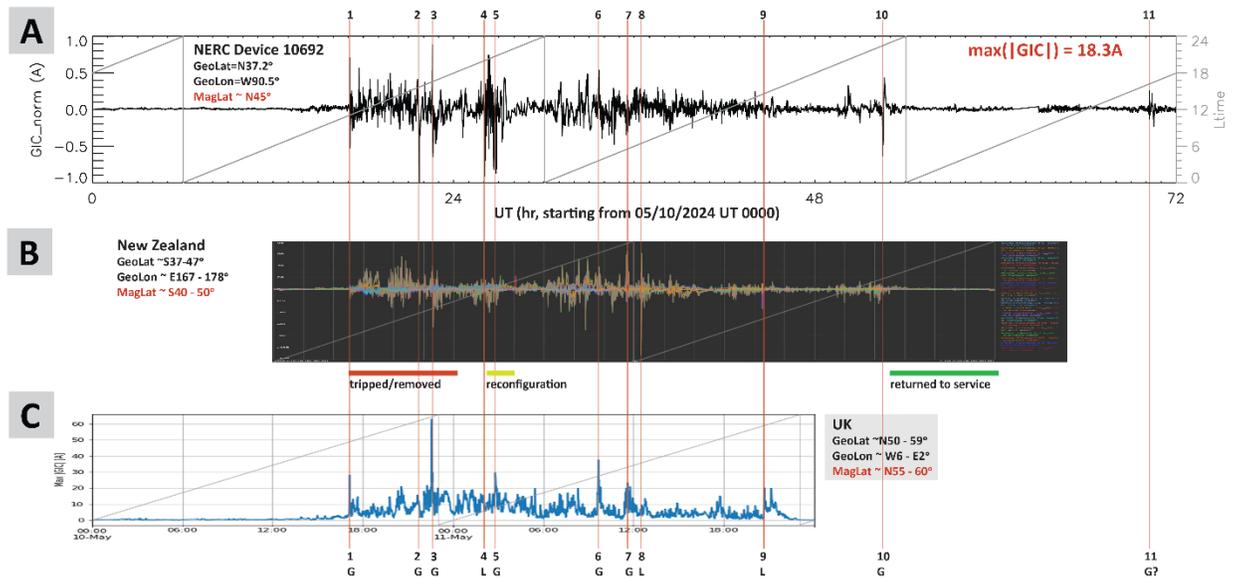

**Figure 6.** Similarities and differences in GICs at geographically distinct locations: Panel A shows the reference place in USA, Panel B shows locations in New Zealand, and Panel C shows locations in the UK. Under Panel B, the color bars roughly indicate the intervals of reported major GIC events, as summarized in the Transpower report. At the bottom, the capital letters "G" and "L" letter below the Time numbers denote moments with "Global" or "Localized" GIC peaks, respectively. Credit: Panel B—Transpower report; Panel C—Lawrence et al. (2025).

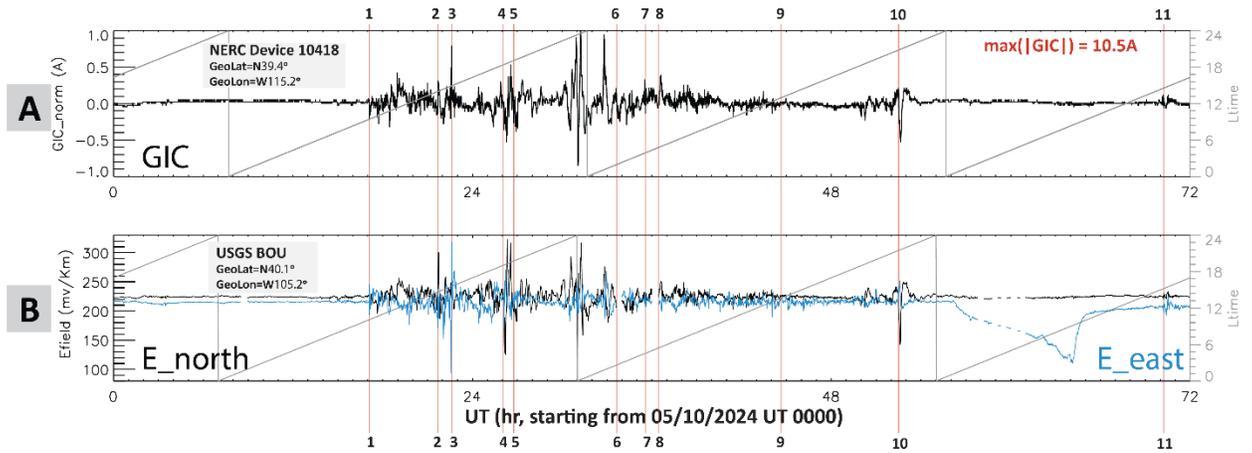

**Figure 7.** Comparison of GIC measurements and induced geoelectric fields. Panel B shows the north-south (in black) and east-west (blue) horizontal components of induced electric field measured at the BOU station (with a geographic latitude 40.1° and longitude of 105.2°), while GIC is recorded by a nearby NERC Device 10418 at a similar latitude (39.4°) and longitude (115.2°).

**Table 1: Timeline of Key Events**

| Date (in 2025) | Time (UTC) | δT (hrs from Solar/L1) | Solar Events | SW Events inside Heliosphere | Effects on Earth — Geomagnetic Disturbances | Effects on Earth — Energetic Particles (RB e-/SEP) | GIC Events |
|---|---|---|---|---|---|---|---|
| May 01 | -- | | AR13664 | | | | |
| May 08 | 0536 | Time zero -- | Multiple large CMEs | | | | |
| May 10 | 1335 | 56.0 -- | | | | Beginning of SEP flux intensification | |
| May 10 | 1637 | 59.0 Time Zero | | IP shocks arrive L1 | | | |
| May 10 | 1705 | 59.5 0.5 | | Forward shock front arrives Earth bow shock nose | | | 1st Global GIC enhancement |
| May 10 | 1707 | 59.5 0.5 | | | SSC and major substorms with AE > 3000 nT | | A list of large GIC spikes |
| May 10 | 1745 | 60.1 1.1 | | | | 1st SPE flux peak | |
| May 10 | … | | | | Storm main phase | Drop-out of MeV electrons | |
| May 11 | 0210 | 68.6 9.6 | | | End of storm main phase; early recovery phase. Major substorms | | Multiple large GIC spikes |
| May 11 | ~0800 | ~71.5 ~12.5 | | | | 2nd SEP flux peak | |
| May 11 | midday | ~76.0 ~17.0 | | | | Start of MeV electron recovery | |
| May 12 | ~0440 | 95.5 26.5 | | Reverse shock front arrives Earth | | | Last major global GIC event during |
| May 12 | ~0600 | ~100.5 ~30.5 | | | Conclusion of major substorms; Start of late recovery phase | | Conclusion of continuous GIC events |
| May 14 | ~0230 | ~141.0 ~81.0 | | | | 3rd SEP flux peak | |
| May 15 | ~0900 | ~172.5 ~112.5 | | | | Maximum of MeV electron fluxes | |